\documentstyle[11pt,aaspp4,psfig]{article}

\font\tenbg=cmmib10 at 10pt

\def \rvecphi{{\hbox{\tenbg\char'036}}}

\begin{document}


\title{ROSSBY WAVE INSTABILITY OF
KEPLERIAN ACCRETION DISKS}

\author{R. V. E. Lovelace}
\affil{Department of Astronomy,
Cornell University, Ithaca, NY 14853;
rvl1@cornell.edu}

\author{H. Li and S.A. Colgate}
\affil{T-6, Los Alamos National Laboratory,
Los Alamos, NM 87545; hli@lanl.gov;
colgate@lanl.gov}

\author{A. F. Nelson}
\affil{Department of Physics, The University of Arizona, Tucson,
AZ 85721; andy@as.arizona.edu}

\begin{abstract}

We find a linear instability of non-axisymmetric Rossby
waves in a thin non-magnetized Keplerian disk
when there is a local maximum in the radial profile
of a key function ${\cal L}(r) \equiv
{\cal F}(r) S^{2/\Gamma}(r)$, where
${\cal F}^{-1} = \hat {\bf z}\cdot ({\bf \nabla}\times {\bf v})
/\Sigma$ is the
potential vorticity, $S = P/\Sigma^\Gamma$ is the entropy,
 $\Sigma$ is the surface
mass density,
$P$ is the vertically integrated pressure,
and $\Gamma$ is the adiabatic index.
  We  consider in detail
the special case where there is a local maximum in the disk entropy
profile $S(r)$.
 This maximum acts to trap the waves in its
vicinity if its height to width ratio
 ${\rm max}(S)/\Delta r$ is larger
than a threshold value.
The pressure gradient derived from
this entropy variation provides the restoring force
for the wave growth.
  We show that the trapped waves act to
transport angular momentum outward.
A plausible way to produce an entropy variation is
when an accretion disk is starting from negligible mass
and temperature, therefore negligible entropy. As
mass accumulates by either tidal torquing, magnetic
torquing, or Roche-lobe overflow, confinement of
heat will lead to an entropy maximum at the outer
boundary of the disk. 
  Possible nonlinear developments from this instability
include the formation of Rossby vortices and the formation of
spiral shocks. What remains to be determined from
hydrodynamic simulations is whether or not
Rossby wave packets (or vortices) ``hold together''
as they propagate radially inward.

\end{abstract}

\section{Introduction}

The central problem of accretion disk theory is how
angular momentum is transported outward so that
matter can accrete onto the central
gravitating object.
Ever since Shakura and Sunyaev (1973)
and Pringle (1981) outlined the
basic structure of accretion
disk (see also \cite{fkr85}),
the origin of enhanced (or $\alpha$)
viscosity has been the subject
of hundreds of research papers.
Classically, high Reynolds number turbulence has
been invoked to explain the enhanced
viscosity.
The driving mechanism(s) of
the turbulence in
high-temperature (for example, Active
Galactic Nuclei) disk systems
has been a long standing
problem because of the {\it apparent} absence of linear
instabilities in non-magnetized
Keplerian disks (see review by \cite{paplin95}).
  For this reason, magnetic fields and
associated instabilities
have been suggested as the origin
of the enhanced viscosity in disks
(\cite{bh98}).
  The widely cited
magneto-rotational instability
(\cite{vel59}; \cite{chan60})
occurs only
when the ${\bf B}$ field is weak.
Also, it
has been studied only for small patches
of a disk using  unphysical boundary
conditions on the magnetic field.
  The nature of the global
angular momentum transport remains
an open problem.

 This paper
reconsiders the question of
the stability of
non-magnetized Keplerian disks.
 In particular, we find a linear
instability of non-axisymmetric
perturbations for conditions where
the disk quantities, such as
surface density and entropy have steep
radial gradients.
 The conditions we consider are
in general nonbarotropic which
distinguish our work from that
of Papaloizou and Pringle (1984, 1985;
Goldreich, Goodman, \& Narayan 1986;
Narayan, Goldreich, \& Goodman 1987).
 Also, in contrast with the work of
Papaloizou and Pringle, the modes
we consider are trapped at least
initially in a narrow
range of radii and therefore do
not depend on reflections from inner
and outer radii of the disk (or tori).

 The instability we find will give rise to
Rossby vortices in the nonlinear limit.
  These vortices are well-known
in thin, two dimensional
planetary atmospheres, for example,
Jupiter's ``Red Spot.''
  The  persistence of
such vortices would be crucial
for the hydrodynamic transport of
angular momentum in accretion disks.
Here, we present our findings of initial
conditions which cause Rossby waves to grow,
and defer the nonlinear behavior of
the vortices to future publications.

We emphasize that the astrophysical relevance
of this instability
is the assumption that in the circumstances where
accretion disks are important, mass at low
temperature and entropy
accumulates  at some external radius to a thickness
such that a condition
of a local maximum in entropy is created
because of confinement of heat.
Then a wave of non-linear Rossby vortex
excitation  carries the mass, and  entropy
maximum  inwards, exciting further vortices
that transport
the angular momentum outwards.

In \S 2, we consider the general stability of a 2D
nonbarotropic disk and point out the role of
a key function ${\cal L}(r)$ in determining the
stability. We then study a case with a significant
`bump' in the radial
distribution of entropy (or temperature) and
calculate its growth rate and dispersion relation.
In \S 3, we discuss the possible
nonlinear evolution of Rossby waves,
especially their role in angular momentum transport.

\section{Theory}

We consider the stability of
a thin (2D), non-magnetized
Keplerian accretion disk.  In treating the
disk as 2D we
assume that the vertical thickness $2h$ is $\ll r$,
perturbation wavelengths are much longer than $2h$.
Using an inertial cylindrical
$(r,\phi,z)$ coordinate system,
we have a surface
mass density $\Sigma(r) = \int_{-h}^{h} dz \rho(r,z)$,
and a vertically integrated
pressure $P(r)= \int_{-h}^{h} dz p(r,z)$,
We consider the equilibrium disk to be steady
($\partial/\partial t=0$) and axisymmetric
($\partial/\partial \phi = 0$),
with the flow velocity
${\bf v} = v_\phi(r) \hat{\rvecphi~}$ (that is,
the accretion velocity $v_r$ is neglected),
the gravitational potential
$\Phi(r)\approx -GM/r$ (the self-gravity is
neglected in this study), and the radial force equilibrium,
$\Sigma v_\phi^2/r = dP/dr
+\Sigma d\Phi/dr$, where
$M$ is the mass of the central object.
The vertical hydrostatic equilibrium gives
$h \approx (c_s/v_\phi)r$, where
$c_s \equiv (\Gamma P/\Sigma)^{1/2}$
is the sound speed and $\Gamma$ is
the adiabatic index.

\subsection{Basic Equations for a Nonbarotropic Disk}

The perturbations considered
are in the plane of the disk
so that the perturbed surface
mass density is
$\tilde{\Sigma} = \Sigma +
\delta \Sigma(r,\phi,t)$;
the perturbed vertically
integrated pressure is
$\tilde{P} = P+\delta P(r,\phi,t)$;
and the perturbed flow
velocity is $\tilde{\bf v} =
{\bf v} +\delta {\bf v}(r,\phi,t)$ with
${\bf \delta v} =
(\delta v_r,\delta v_\phi,0)$.
The equations for the
perturbed disk are
$$
{D \tilde{\Sigma}\over Dt}
+ \tilde{\Sigma}~ {\bf \nabla}\cdot
\tilde{\bf v} = 0~,
\eqno(1a)
$$
$${D \tilde{\bf v}\over Dt}  =
-{1\over \tilde{\Sigma}}
{\bf \nabla} \tilde{P} - {\bf \nabla}\Phi ~,
\eqno(1b)
$$
$$ {D \over Dt} \bigg
[{\tilde{P} \over \tilde{\Sigma}^\Gamma}\bigg]
= 0~,
\eqno(1c)
$$
where $D/Dt \equiv \partial /\partial t
+ {\bf v}\cdot {\bf \nabla}$ and
we refer to $S \equiv P/\Sigma^\Gamma$
as the entropy of the disk matter.
Equation (1c) corresponds to the isentropic behavior of
the disk matter.  We use equation (1c) for
simplicity rather than
the rigorous relation $D(p/\rho^\gamma)/Dt = 0$
(see Papaloizou and Lin 1995).

We consider perturbations $\propto f(r){\rm exp}
(im\phi - i \omega t)$, where $m$ is the azimuthal
mode number and $\omega$ the angular frequency.
 From equation (1a), we have
$$i\Delta \omega~ \delta \Sigma =
{\bf \nabla} \cdot
(\Sigma ~\delta {\bf v})~,
\eqno(2)
$$
where
$\Delta \omega(r) \equiv \omega - m \Omega(r)$ and
$\Omega = v_\phi/r \approx (GM/r^3)^{1\over 2}$.
 From equation (1b) we have
$$ i\Delta \omega \delta v_r +
2 \Omega \delta v_\phi =
{1\over \Sigma}{\partial \delta P \over \partial r}
-{\delta \Sigma \over \Sigma^2} {\partial P
\over \partial r}~,
\eqno(3a)
$$
$$
i\Delta \omega \delta v_\phi -
{\kappa^2 \over 2 \Omega}
\delta v_r = ik_\phi
 {\delta P \over \Sigma}~.~~~~~~~~~~~~~~
\eqno(3b)
$$
Here,
$\kappa \equiv
[r^{-3} d(r^4 \Omega^2)/dr]^{1\over 2}$
$\approx \Omega$ is the radial
epicyclic frequency and $k_\phi = m/r$
is the azimuthal wavenumber.
 From equation (1c), we have
$$\delta P = c_s^2\delta \Sigma -
{ i\Sigma c_s^2 \over \Delta \omega L_s} \delta v_r~,
\eqno(4)
$$
which is the perturbed equation of state and
$$ L_s \equiv \Gamma \bigg/
\bigg[{d \over dr}~{\rm ln}
\bigg( {P\over \Sigma^\Gamma }\bigg) \bigg]~
\eqno(5)
$$
is the length scale of the entropy
variation in the disk. The nonbarotropic nature
of the disk renders $L_s$ to be finite whereas
for a barotropic disk (i.e., $P$ is a function
of $\Sigma$ only), $L_s \rightarrow \infty$.

To further simplify the momentum equation,
we can write the right-hand-side of
equation (3a) as $(\delta P/\Sigma)^\prime
+(\delta P/\Sigma^2)\Sigma^\prime -
(\delta \Sigma /\Sigma^2) P^\prime$,
where the prime denotes
$\partial/\partial r$.
Next, equation (4) can be used to
express this as $(\delta P/\Sigma)^\prime
-\delta P/(\Sigma L_s) -i\delta v_r c_s^2/
(\Delta \omega L_s L_p)$, where
$$
L_p \equiv \Gamma \bigg/\left[
{d \over dr}~ {\rm ln}(P)\right]~
\eqno(6)
$$
is the length-scale of the pressure
variation.
 For definiteness we assume
$h<(|L_s|,|L_p|) {\buildrel < \over \sim}~ r $.

Using $\Psi \equiv \delta P/\Sigma$ as our key
variable,
we can now rewrite equations (3) as
$$
i\left(\Delta \omega +{c_s^2 \over
\Delta \omega L_s L_p}\right)\delta v_r +
2 \Omega \delta v_\phi =
{\partial \Psi\over \partial r}-{\Psi \over L_s},
\eqno(7a)
$$
$$
~~~~~~~~~~i\Delta \omega \delta v_\phi -
{\kappa^2 \over 2 \Omega}
\delta v_r = ik_\phi \Psi~.
\eqno(7b)
$$
Solving equations (7) for $(\delta v_r,\delta v_\phi)$
gives
$$ \Sigma~ \delta v_r = i~{\cal F}
\left[{\Delta \omega \over \Omega}\left(\Psi^\prime-
{\Psi \over L_s}\right)
 -2k_\phi \Psi \right]~,
\eqno(8a)
$$
$$
\Sigma ~\delta v_\phi = {\cal F} \bigg[
-k_\phi\left({\Delta \omega \over \Omega} + {c_s^2 \over
\Delta \omega \Omega L_sL_p}\right)\Psi +~~~~
$$
$$
+{\kappa^2\over 2 \Omega^2}\left(\Psi^\prime-
{\Psi\over L_s}\right) \bigg]~,
\eqno(8b)
$$
where
$${\cal F}(r)\equiv {\Sigma ~\Omega \over
\kappa^2 - \Delta \omega^2 -c_s^2 /( L_sL_p)}~.
\eqno(9)
$$

For `corotation modes' where
$|\Delta \omega|^2$  is small
compared with $\kappa^2$ and $c_s^2/|L_sL_p|$,
we have
${\cal F} \approx \Sigma \Omega/\kappa^2
= \Sigma/2\omega_z$.
Here, $\omega_z \equiv
\hat {\bf z}\cdot ({\bf \nabla}\times {\bf v})$
is the vorticity.  The ratio $\omega_z/\Sigma
\approx 1/(2\cal{F})$
is commonly referred to as the
potential vorticity (see for example
Drazin 1978).

We next use the continuity equation (2) and
equations (4) and (8) to obtain
$${1\over r}\left({r {\cal F} \over \Omega}
\Psi^\prime\right)^\prime
-{k_\phi^2{\cal F}\over \Omega} \Psi
={\Sigma \Psi \over c_s^2}
+ {2k_\phi {\cal F}^\prime\over \Delta \omega}\Psi
$$
$$+\left[{{\cal F} \over \Omega L_s^2}+
{1\over r}\left({r{\cal F}\over \Omega L_s}\right)^\prime
+{ 4 k_\phi {\cal F} \over \Delta \omega L_s}+
{k_\phi^2 c_s^2 {\cal F} \over
  \Delta \omega^2 \Omega L_sL_p} \right]\Psi~.
\eqno(10)
$$

Equation (10) allows the determination
of $\omega$ and identification
of mode structure for  general disk flows,
both stable or unstable.
If the flow is barotropic (as
has been studied extensively in the literature), 
all the terms on the lower line of this
equation vanish,  $|L_s| \rightarrow \infty$.
In this paper, we explore additional effects 
which arise when
$|L_s|$ is finite.

\subsection{Instability Condition}

A useful quadratic form can be obtained
by multiplying equation (10) by $\Psi^*$ and
integrating over the disk.
Assuming $\Psi^*\Psi^\prime r
{\cal F}/\Omega \rightarrow 0$
for $r\rightarrow 0,~\infty$, we obtain
$$-\int d^2r~{{\cal F}\over \Omega}
\left(|\Psi^\prime|^2 + k_\phi^2 |\Psi|^2\right)
=\int d^2r~ {\Sigma \over c_s^2} |\Psi|^2~+
$$
$$
\int d^2r~ \left[ {{\cal F} \over \Omega L_s^2}
+{1\over r}\left( { r{\cal F}
\over\Omega L_s}\right)^\prime\right] |\Psi|^2~+
$$
$$
2\int d^2r~\left({k_\phi {\cal F}^\prime
\over \Delta \omega}+{2 k_\phi {\cal F}
\over \Delta \omega L_s}\right) |\Psi|^2~+
$$
$$
\int d^2r~ {k_\phi^2 c_s^2 {\cal F}
\over \Delta \omega^2 \Omega L_sL_p} |\Psi|^2~.
\eqno(11)
$$
Note that only the last
two integral terms on the right-hand side of
this equation involve the
complex variable $\Delta \omega(r)$.

Equation (10) can be solved by several methods
(e.g., by a relaxation method or by shooting), but we defer the full
discussion of its solutions to a future publication.
In order to bring out the essential aspects of equation (10),
we make simplifying assumptions which allow an analytic solution. 
First, when
$|\Delta \omega|$ is a finite fraction of $\Omega$
and $c_s/v_\phi \ll 1$, we find that
the integral involving $\Delta \omega^2$
is small compared to the
integral with $\Delta \omega$, so that
the last term in equation (11) is negligible.
Furthermore, for the `corotation modes' where
$|\Delta \omega|^2/\Omega^2 \ll 1$, 
and for conditions where the pressure variation with $r$
is such that $\kappa^2/\Omega^2$ is positive,
the function ${\cal F}$ is positive and real. It is physically
possible to have $\kappa^2$ positive and yet have
a finite $|L_s|$ because the pressure and entropy
variations need not to be the same. In fact, this is precisely
a feature of nonbarotropic flow (see next section). 
Thus, the terms containing $\Delta \omega$ in (11) can
be combined to give
$$
2\int d^2r~{k_\phi {\cal F} \over \Delta \omega}
\left[ {\rm ln} \left( {\cal F} S^{2/\Gamma} \right)\right]^\prime
|\Psi|^2~.
\eqno(12)
$$

Putting equation (12) back into (11) and
taking the imaginary part of (11), we find that
a solution is possible only for conditions where the imaginary
part of (12) vanishes (see also Lovelace \& Hohlfeld 1978; 
hereafter LH).
Thus, we find that an
instability is possible only if 
$\left[ {\rm ln} \left( {\cal F} S^{2/\Gamma} \right)\right]^\prime$
vanishes at some $r$, in other words,
the key function
${\cal L}(r) \equiv {\cal F}(r) S^{2/\Gamma}(r)$
has a maximum or minimum.

This newly defined function ${\cal L}(r)$
warrants further discussion. We believe that this
expression is new and more general than those given
in previous studies. With the above assumptions,
we can write  ${\cal L}(r)$ as
$$
{\cal L}(r) = {\cal F} S^{2/\Gamma}
\approx \frac{\Sigma ~\Omega~ S^{2/\Gamma}}{\kappa^2}
\propto \frac{\Omega}{\kappa^2}~ \Sigma^{2/\Gamma -1}
T^{2/\Gamma}~,
\eqno(13)
$$
where $\Omega^2(r)$ and $\kappa^2(r)$
are of course constrained by the
radial force balance in the equilibrium.

An extremum of
${\cal L}(r)$ necessary for instability
could come from several
sources: (a) In the limit
of incompressible and isothermal flow
(cf. \cite{sg79} for an application to
flow in a tornado);
 an extremum in the radial profile
of $\Omega(r)$ is required
to have  unstable vortices.
(b) In the studies of LH and
Sellwood \& Kahn (1991),
a cold disk was assumed and
$S(r)$ was effectively
taken as a constant.
In this case
${\cal L}(r)$ is simply ${\cal F}
= \Sigma/\Omega$, which
must have a maximum or a minimum for instability.
  Similarly, in the  cases considered by
Papaloizou \& Pringle (1984, 1985),
where a polytropic (barotropic)
$P=P(\Sigma)$ equation of
state is assumed,
$S(r)$ is again a constant
throughout the disk so that
a maximum or minimum of
${\cal L}(r) \propto {\cal F}$
is needed for instability.
(c) In the present study,
we treat the more general case
where $S(r)$  varies across
the disk.
  An extremum in ${\cal L}(r)$
can result from
having an extremum in $S(r)$
directly (that is, a local
 ``bump'' in the temperature
profile of the disk), or by having a large
enough gradient  in $S(r)$
which causes an extremum in the
epicyclic frequency
$\kappa^2$ owing to the
radial force balance.
  Furthermore, we find below that
instability occurs when
a threshold in the
variation of ${\cal L}(r)$ is exceeded.
In view of equation (13),
a variation of $T(r)$ is more important
in giving a change in ${\cal L}(r)$
than a comparable fractional variation of
$\Sigma(r)$.

\subsection{Thermodynamic Driving of Vorticity}

It is straightforward to obtain the vorticity equation
(cf. \cite{ped87}; \cite{spi93}) by taking the curl of
equation (1b) and combining with (1a),
$$
\frac{D  }{Dt}~\left({ \omega_z \over \Sigma}\right) =
\frac{{\bf \nabla}\Sigma \times {\bf \nabla}P }{\Sigma^3}~,
\eqno(14)
$$
where again, ${\omega}_z=\hat{\bf z}\cdot
{\bf \nabla}\times {\bf v}$ is the
vorticity.
The right-hand side of this equation
represents the
thermodynamic driving  of the vorticity.
For a barotropic flow, $P=P(\Sigma)$,
$$
\frac{D  }{Dt}~\left({\omega_z\over\Sigma}\right) = 0~.
\eqno(15)
$$
This is the case for most
previous studies (cf. \cite{pj84};
\cite{pj85}; \cite{gol86}; \cite{nar87}).
Equation (15) says
that each fluid element
conserves its  specific vorticity.
 In contrast, the  term ${\bf \nabla}\Sigma
\times {\bf \nabla}P
\propto {\bf \nabla}T \times {\bf \nabla}S$
 destroys this conservation and allows
pressure forces to ``produce''
vorticity in the flow.
 In the non-barotropic case
surfaces (or lines) of constant surface density
and surfaces of constant specific entropy
do not coincide.

\subsection{Dispersion Relation}

For the `corotation modes'
($|\Delta \omega|^2 \ll \Omega^2$)
and ${\cal F}(r) \approx \Sigma/\Omega$,
an approximate dispersion
relation follows from equation (10)
by taking the $r-$dependence
$\Psi \propto {\rm exp}(ik_r r)$ with
$(k_r r)^2 \gg 1$, $(k_rL_s)^2 \gg 1$,
and $(k_rL_p)^2 \gg 1$.
The dominant terms
in (10) are:
$$-{{\bf k}^2 \Sigma \over \Omega^2} \Psi =
{\Sigma \over c_s^2} \Psi+{2k_\phi \Sigma \over
\Delta \omega \Omega}
\left( {\rm ln}{\cal L}\right)^\prime \Psi
+{ k_\phi^2 c_s^2 \Sigma \over
\Delta \omega^2 \Omega^2 L_sL_p}\Psi~,
\eqno(16)
$$
or
$$
\Delta \omega =-{k_\phi c_s^2/\Omega
\over 1+{\bf k}^2 h^2 }
\left[({\rm ln}{\cal L})^\prime
\pm {\sqrt{ [({\rm ln}{\cal L})^{\prime}]^2
-{1+{\bf k}^2 h^2 \over L_sL_p}}}~\right]~,
\eqno(17)
$$
where ${\bf k}^2 \equiv k_r^2+k_\phi^2$ and
$h \equiv c_s/\Omega$.
The dispersion relation (17) is the
disk analogue of the Rossby wave dispersion
relation (see for example
Brekhovskikh \& Goncharov 1993).
Note in particular that the group velocity
of the waves in the corotating reference
frame is in the direction opposite
to that of the disk rotation.
Instability is possible if
$[(\ln{\cal L})^{\prime}] ^2
< (1+{\bf k}^2 h^2)/(L_s L_p)$,
which requires $L_sL_p>0$.

From equation (17), the maximum growth
rate occurs for $(\ln {\cal L})^\prime =0$,
and it is
$${{\rm max}(\omega_i)\over \Omega}=
{|k_\phi|c_s^2/\Omega^2 \over
(1+{\bf k}^2h^2)^{1\over 2} (L_sL_p)^{1\over2}}~
$$
$$
\approx {|k_\phi|h \over (1+{\bf k}^2h^2)^{1\over 2}}
\left({c_s \over v_K}\right)
\left( {r^2 \over L_sL_p} \right)^{1\over 2}~,
\eqno(18)
$$
where $v_K$ is the Keplerian
velocity and $\omega_i \equiv {\cal I}m(\omega)$
is the growth rate.
The maximum growth rate is $< \Omega$,
for a thin disk where $h/r
\approx c_s/v_{\phi} \ll 1$ and for $L_sL_p > h^2$.
 Note in particular the geometrical optics
relation for a wavepacket,
$$
{ d k_r \over dt} = -{\partial \omega
\over \partial r}
\approx {3\over 2} k_\phi \Omega~,
~{d k_\phi \over dt}=-{\partial \omega \over
\partial \phi} = 0~,
\eqno(19)
$$
so that $ k_r \approx {3\over 2} k_\phi \Omega t
+ $const., and $k_\phi =$ const.
 Thus, a wave packet initially
with $|k_rh| \ll1$ is rapidly sheared by
the differential rotation.
 In a time
of order $t_{shear} =
({3\over 2} k_\phi h \Omega)^{-1}$ the wave
evolves into having a very short radial
wavelength,  $|k_r h| >1$, trailing spiral
wave.
 This is discussed in detail by Goldreich
and Lynden-Bell (1965).  The maximum
amplification a wave can have is
$A = {\rm exp}(\int dt~ \omega_i)\sim
{\rm exp}[{\rm max}(\omega_i) t_{shear}]
\approx {\rm exp}
[ (2/3)(c_s/v_K)(r^2/L_sL_p)^{1\over2}]$.
The amplification is $A={\cal O}(1)$ and is
probably not significant
for $L_sL_p > h^2$.

\subsection{Trapped Modes}

Here, we consider conditions where
${\cal L} \approx
(\Sigma / \Omega) S^{2/\Gamma}$
has a  maximum or minimum as a function of
$r$ at $r_0$ of width $\Delta r \ll r_0$.
 The case where ${\cal F}(r) = \Sigma/\Omega$
has an
extremum  was treated earlier by LH.
 Therefore, we consider the case where
${\cal F} \approx$ const and where $S(r)$ has an
extremum.
 The extremum of $S(r)$ is thus an extremum
of $T(r)$.
 As mentioned earlier, a variation
of $T(r)$ is more important in giving a
change in ${\cal L}(r)$
than the variation of $\Sigma(r)$.
 For a maximum in $S(r)$,
we find using equations (17)
and (19) that the radial group velocity
of the waves $v_{gr} =
\partial \omega/\partial k_r$
is directed towards $r_0$, whereas
for a minimum the waves propagate away
from $r_0$.  Thus, a maximum of
$S(r)$ can give
trapping of the waves in
the vicinity of $r_0$.
 This
trapping obviates the need
for reflecting disk boundaries considered
by Goldreich et al. (1986).

For definiteness we take
$$
S(r) = S_1 +(S_2-S_1){\rm exp}
[-(r-r_0)^2/(2 \Delta r^2)]~,
\eqno(20)
$$
where $S_1, S_2, r_0,$ and $\Delta r$
are constants.
For simplicity we consider
$\Delta r^2 \ll r_0^2$.
 In this limit, the real and
imaginary parts of equation (10)
for $\Psi = \delta P/\Sigma = \Psi_r + i \Psi_i$ can
be written as
$$ \Psi_r^{\prime \prime}=(K^2)_r~\Psi_r - (K^2)_i~\Psi_i~,
$$
$$  \Psi_i^{\prime \prime}=(K^2)_r~\Psi_i + (K^2)_i~\Psi_r~.
\eqno(21)
$$
Here, we have omitted the detailed expressions for
$(K^2)_r$ and $(K^2)_i$, both of which are real functions
of $L_s$, $c_s/v_{\phi}$, $\Delta \omega/\Omega$ and $m$
(see figures).
It can be shown that $(K^2)_r$ and $(K^2)_i$ possess the
following properties (assuming $\omega_r/[m\Omega(r_0)] =1$):
$(K^2)_i$ is an odd function
of $r-r_0$ and is $\rightarrow 0$ when $(r-r_0)^2 \gg \Delta r^2$;
but $(K^2)_r$ is an even function of $r-r_0$.
$(K^2)_r$ is $< 0$ for $r \rightarrow r_0$, which corresponds to
a classically allowed region of motion in a potential well,
and is $> 0$ for $(r-r_0)^2 \gg \Delta r^2$, which corresponds to a
forbidden region. It is this potential well give rise to
``bound states'' which determines
the growth rates $\omega_i$.

The existence of wave growth ($\omega_i >0$) and its dependence
on $S_2/S_1$ can be obtained by first considering
the potential well for $\omega_i \rightarrow 0$.
The first appearance of a bound state in the well is
given by the Bohr-Somerfeld quantization condition,
$$
f(S_2/S_1) \equiv \oint dr(-K^2)_r^{1\over2} = \pi~.
\eqno(22)
$$
This condition corresponds to a critical
value of $S_2/S_1$ so that
there can be a bound state with $\omega_i/\Omega >0$
(which is the middle curve in Figure 1).
This is physically the threshold required
for the growth of Rossby waves.
Again, we emphasize that the existence of
wave growth depends critically on the nonbarotropic
nature of disk, because the main contribution to
this potential well comes from  $4k_\phi{\cal F}\Psi
/(\Delta \omega L_s)$, which is always nonpositive
since ${\cal R}e(1/\Delta \omega)
\propto r-r_0$ whereas $1/L_s \propto r_0-r$.
This term vanishes when $L_s \rightarrow \infty$.

\begin{figure}
\psfig{file=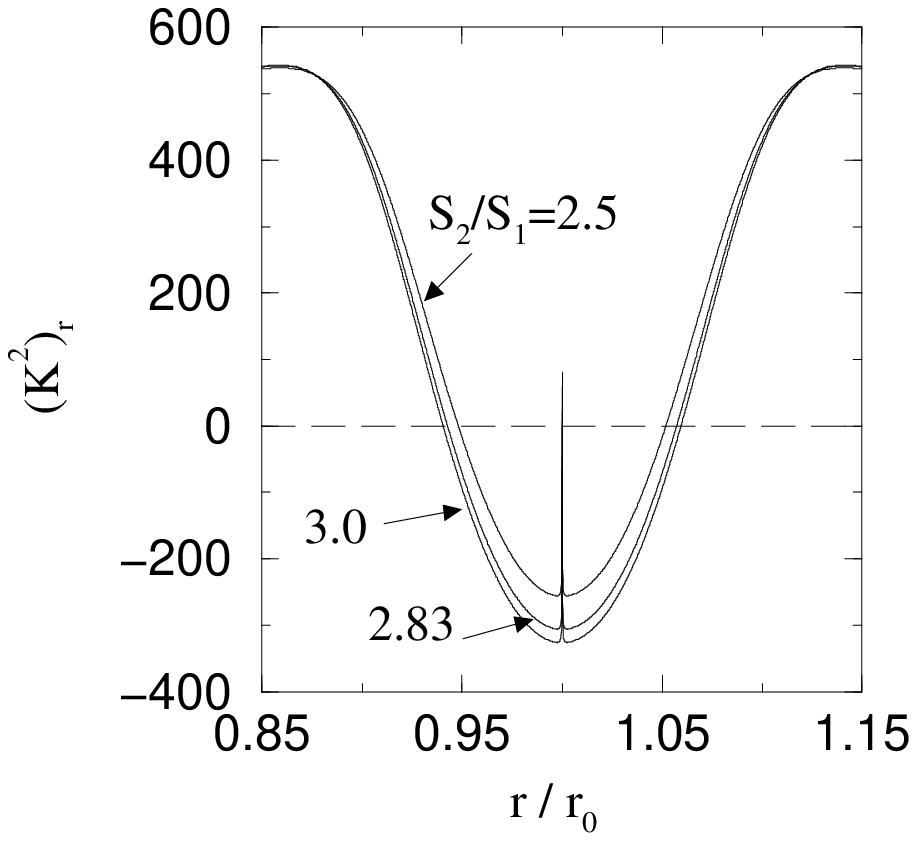,width=6in,height=5in,angle=0}
\caption{
Sample profiles of $(K^2)_r$
in units of $1/r_0^{2}$
are shown
for different values of $S_2/S_1$.  For this
plot $m=5$, $\Delta r/r_0=0.05$, $\Gamma=5/3$,
unperturbed $c_s/v_{\phi}=0.045$, $L_s = L_p$,
and $\omega_i/\Omega=10^{-3}$.
The spike at $r/r_0=1$ is due to the
finite value of $\omega_i$.
For the intermediate curve,
the Bohr-Somerfeld condition (10)
is satisfied to a good approximation.
For other values of $m$
between $1$ and $10$, the critical
value for the bound state is $S_2/S_1 \approx
2.7+0.83(m/10)^{5\over2}$.
\label{fig1}}
\end{figure}

\placefigure{fig1}

Figure 2 shows sample
results for $(K^2)_{r,i}$ and $\Psi_{r,i}$ by
solving equations (21) using a shooting method.
We omit discussing the detailed numerical method,
but only point out that one can make use of
the symmetry of the potential well as well as
the boundary conditions at $r=r_0$ and
$r \rightarrow \infty$.
The key physical variables are
$S_2/S_1$ and $\omega_i/\Omega$.
We find $\omega_i$ to have the approximate dependence
${\omega_i / \Omega} \approx
k_1[ (S_2 /S_1) -k_2]^{2 / 3}$
for $S_2/S_1 \geq k_2$, where $k_2$
is the mentioned
critical value and $k_1$ is a constant.

\begin{figure}
\psfig{file=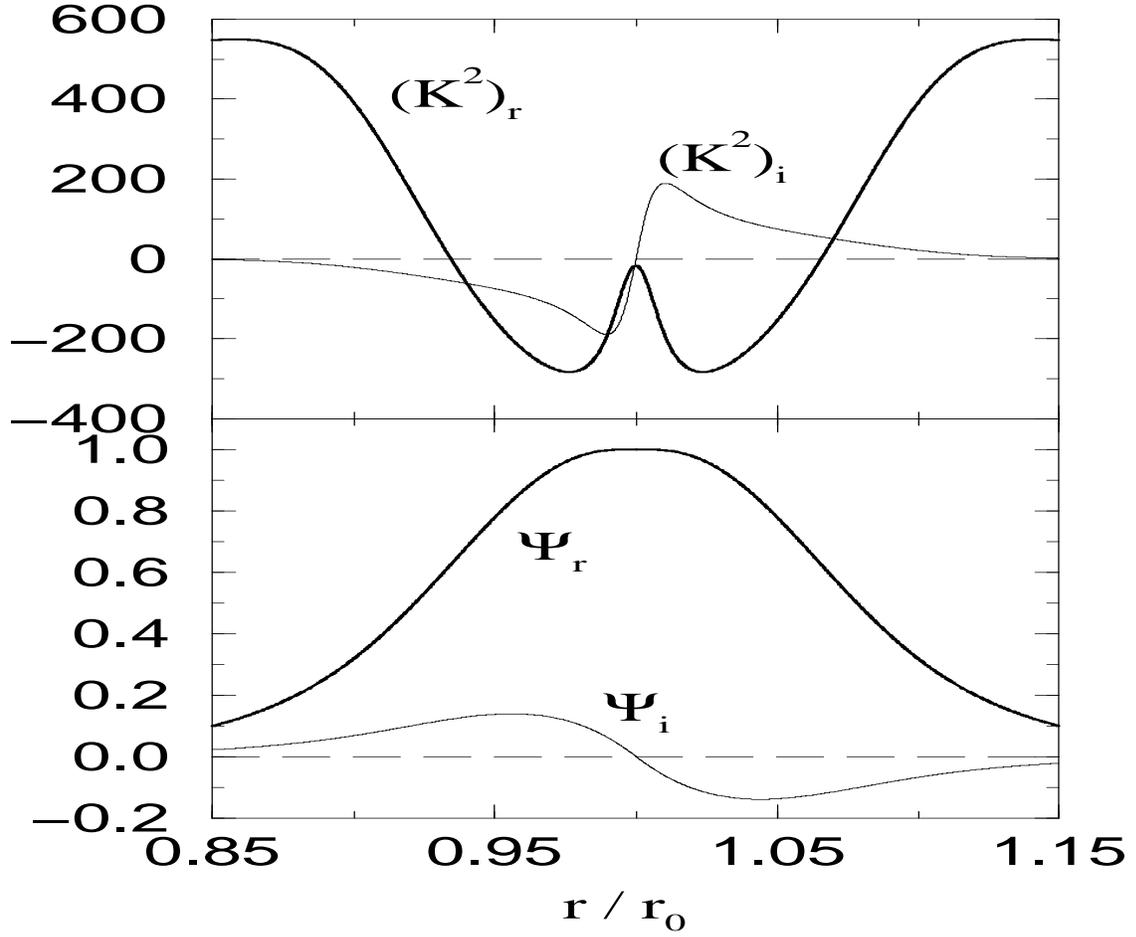,width=6in,height=5in,angle=0}
\caption{
Sample profiles of $(K^2)_{r,i}$
and $\Psi_{r,i}$, obtained by a shooting
method,
are shown for $\omega_i/\Omega=0.1$, $m=5$,
and $\Delta r/r_0=0.05$.  For this case,
$S_2/S_1 \approx 3.744$ and
$\Psi^\prime_i(r_0)
\approx -6.543$.
\label{fig2}}
\end{figure}

\placefigure{fig2}

Figure 3 shows the vortical nature of the
velocity field for the same unstable case as Figure 2.

\begin{figure}
\psfig{file=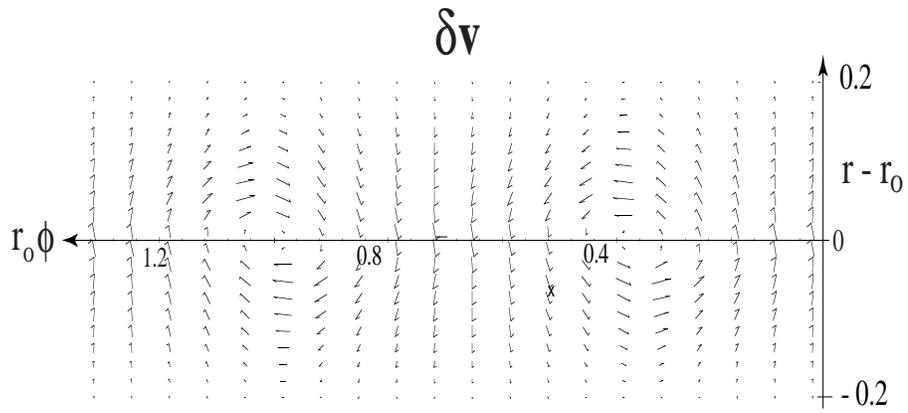,width=6in,height=5in,angle=0}
\caption{Velocity field of the
unstable perturbation shown in
Figure 2, obtained from equations (8)
and (21).
\label{fig3}}
\end{figure}

\placefigure{fig3}

\subsection{Angular Momentum Transport}

To calculate the  flux of angular momentum,
we first note that, with the complex factors
$\exp(im\phi -i\omega t)$ removed,
$\delta v_r =\delta v_r^O +i\delta v_r^E$,
$
\delta v_\phi =\delta v_\phi^O +i\delta v_\phi^E$
and
$
\delta \Sigma =\delta \Sigma^E +i\delta \Sigma^O
$,
where the ${E,O}$ superscripts denote even
or odd {\it real} functions of $r-r_0$.
Such properties are obtained by expressing
them using equation (1) and $\Psi = \Psi_r^E + i \Psi_i^O$.
Thus across a circle of radius $r_0$,
we have
$$F_J(r_0) = \int_0^{2\pi} d\phi~
r_0^2~ {\cal R}e(\Sigma) {\cal R}e(\delta v_r)
{\cal R}e(v_{\phi})~,
$$
$$
= \pi r_0^2~ \Sigma~ \delta v_r^E(r_0)~
\delta v_\phi^E(r_0)~,
$$
$$={2 \pi m^2 \Sigma \omega_i \over \Omega^3}
\left(\Psi_r +{\omega_i r_0 \over 2\Omega m}
\Psi_i^\prime \right)_{r_0}
\left( \Psi_r - {\Omega r_0 \over 2 \omega_i m}
\Psi^\prime_i \right)_{r_0}.
\eqno(23)
$$
We find in general that $m \Psi^\prime_i(r_0) <0$.
Further, we find that each of the two factors
in parentheses in (23) are positive.  Therefore,
the trapped wave causes the
{\it outward} flow of angular
momentum across $r=r_0$.
Furthermore, one notes that the positive transport
is from a ``nonlinear'' term ($\delta v_r \delta v_\phi$)
since the first order term vanishes.

\section{Discussion}

The astrophysical importance of this Rossby wave
instability is the implication
that in the nonlinear limit,
multiple Rossby vortices will form after
these waves break and coalesce in a few revolutions.
Hence, they provide a plausible
explanation of the hydrodynamic
transport of angular momentum
in thin, Keplerian  accretion disks.
The critical condition for
the inception of this hydrodynamic
transport, as derived in this
linear case, is a minimum
contrast ($\sim \times 3$) in
the radial distribution of
entropy. Thus, it is important
that heat (and thus entropy)
be preserved against cooling by
radiation in at least tens of revolutions.
Roughly speaking, recognizing
the large dynamic range and uncertainty
of opacities, this implies
a minimum mass `thickness' $\Sigma \sim
100-1000$ g cm$^{-2}$.
An important implication from this
minimum mass `thickness' requirement
is the prediction on the
turn-on (active state) and turn-off (quiescent state)
conditions for the angular
momentum transport
by the Rossby vortices.
Specifically, we could envisage the following scenario:
as mass accumulates in the disk by accretion, the
resulting trapping of heat allows the
temperature to rise by the
energy input of accretion and
an episode of Rossby vortex
accretion ensues.  Depending
upon the heating, opacity
conditions, and mass accretion rates,
there will be a radius, a
thickness, and therefore a
mass  where the Rossby instability
initiates.  This mass
and mass accretion rate
determine the event luminosity and
episode rate.  Finally we note that a given episode
terminates when its mass
is accreted presumably rapidly
before additional mass is supplied.
This understanding
of the turn-on and turn-off conditions for Rossby vortex
transport may prove relevant for several astrophysical
systems, such as the planet formation
after the solar
nebula has cooled (Cameron 1962) and outbursts from CVs
and stellar mass black holes (Vishniac 1997).

Although a localized high
entropy region ($S_2/S_1 \sim 3$)
may appear to be a large deviation in $S(r)$,
the resultant pressure gradient force
is still very small compared
to the gravitational force. In the case represented
by all the figures, we have used an unperturbed
$c_s/v_{\phi}=0.045$, which gives $c_s/v_{\phi} \approx 0.078$
at the peak of perturbed region ($S_2/S_1 \sim 3$).
 But this deviation in $S(r)$ is sufficient to
drive the growth of non-axisymmetric perturbations.
 A possible mechanism of formation of an entropy bump
is from the interaction between the matter
from Roche lobe overflow and the accretion disk.
 A spiral shock can form in this situation  (Sawada et al. 1986).

 Bracco et al. (1998) have
studied {\em barotropic} disks
with initial perturbations in the potential vorticity
($\omega_z/\Sigma$) profile. They showed that vortices
can have a rather long life time.
  Another possible nonlinear effect is the formation of
spiral shocks which heat the disk and thereby maintain the bump
against radiative cooling.
  In either case, requiring a finite entropy bump
for the instability to grow implies a maximum
cooling rate above which a bump which
develops in the disk quickly decays away and
such an enhanced viscosity model turns off.
   What remains to be determined
is whether the bump ``holds together'' as it
propagates radially inward.
  If these Rossby vortices are indeed
playing a major role in
angular momentum transport, then
they are fundamentally
different from the usual $\alpha$
viscosity prescription which
is based on homogeneous turbulence.
  In contrast, Rossby vortices
are large scale, coherent structures.

Preliminary numerical simulations using a
high resolution PPM code
(Nelson et al. 1998) appear
to confirm the existence of
a threshold on $S_2/S_1$ for wave growth
and the formation of anti-cyclonic vortices.
Numerical studies are now underway
to follow the nonlinear evolution
of the vortices in nonbarotropic
disks and to study the
associated
transport properties (Nelson et al. 1998).

\acknowledgements{
We thank the referee's insistence in clarifying
many of issues which greatly improved this paper.  
The experimental work on the flow visualization
for a liquid sodium dynamo by Ragnar Ferrel and
Dr. Van Romero is gratefully acknowledged in
stimulating this analysis. Discussion with
P. Yecko is also gratefully acknowledged.
The work of RVEL was supported in part
by NSF grant AST-9320068 and NASA grant
NAG5-6311.
 HL gratefully acknowledges the support
of an Oppenheimer Fellowship.
 The work of HL and SAC
was performed under the
auspices of the U.S. Department of Energy.
}

\newpage

\end{document}